\documentclass[letterpaper,conference]{IEEEtran}

\usepackage{graphicx}
\usepackage[utf8]{inputenc} 
\usepackage[T1]{fontenc}
\usepackage{url}
\usepackage{ifthen}
\usepackage{cite}
\usepackage{color}

\usepackage{cite}

\ifCLASSINFOpdf
\else
\fi
\usepackage{array}
\usepackage{graphicx}
\usepackage{hyperref}
\usepackage{amsmath}
\usepackage{amssymb}
\usepackage{bm}
\usepackage{bbm}
\usepackage{algorithmicx}
\usepackage{algorithm}
\usepackage{algpseudocode}
\usepackage{array}
\usepackage{booktabs}
\usepackage{multirow}
\usepackage{makecell}
\usepackage{tikz}
\usetikzlibrary{shapes.geometric, arrows}

\begin{document}
%
\title{Design, Performance, and Complexity of CRC-Aided List Decoding of Convolutional and Polar Codes for Short Messages}

\author{Jacob King\IEEEauthorrefmark{1}\IEEEauthorrefmark{3},
Hanwen Yao\IEEEauthorrefmark{2},
William Ryan\IEEEauthorrefmark{3},
and
Richard D. Wesel\IEEEauthorrefmark{1},\\

\IEEEauthorblockA{\IEEEauthorrefmark{1}Department of Electrical and Computer Engineering, University of California, Los Angeles, Los Angeles, CA 90095, USA} 

\IEEEauthorblockA{\IEEEauthorrefmark{2}Department of Electrical and Computer Engineering, University of California, San Diego, San Diego, CA 92093, USA} 

\IEEEauthorblockA{\IEEEauthorrefmark{3}Zeta Associates, Aurora, Denver, CO 80011, USA} 

Email: jacob.king@ucla.edu, hay125@eng.ucsd.edu, ryan-william@zai.com, wesel@ucla.edu
}


\maketitle

\begin{abstract}

Motivated by the need to communicate short control messages in 5G and beyond, this paper carefully designs codes for cyclic redundancy check (CRC)-aided list decoding of tail-biting convolutional codes (TBCCs) and polar codes.  Both codes send a 32-bit message using an 11-bit CRC and 512 transmitted bits.  We aim to provide a careful, fair comparison of the error performance and decoding complexity of polar and TBCC techniques for a specific case.  Specifically, a TBCC is designed to match the rate of a (512, 43) polar code, and optimal 11-bit CRCs for both codes are designed.  The paper examines the distance spectra of the polar and TBCC codes, illuminating the different distance structures for the two code types.   We consider both adaptive and non-adaptive CRC-aided list decoding schemes.  For polar codes, an adaptive decoder must start with a larger list size to avoid an error floor.  For rate-32/512 codes with an 11-bit CRC, the optimized CRC-TBCC design achieves a lower total failure rate than the optimized CRC-polar design. Simulations showed that the optimized CRC-TBCC design achieved significantly higher throughput than the optimized CRC-polar design, so that the TBCC solution achieved a lower total failure rate while requiring less computational complexity.
\end{abstract}



{\let\thefootnote\relax\footnote{{This research is supported by Zeta Associates Inc. and National Science Foundation (NSF) grant CCF-2008918. Any opinions, findings, and conclusions or recommendations expressed in this material are those of the author(s) and do not necessarily reflect views of Zeta Associates Inc. or NSF.}}}

%
\IEEEpeerreviewmaketitle

\section{Introduction}
\label{sec:Intro}

Polar codes are a class of capacity-achieving codes first introduced by Ar{\i}kan in \cite{ArikanPolar}.  Since their introduction, polar codes have seen wide interest and application.  One such application is in the 5G standard, for use on the physical broadcast channel (PBCH).  In the 5G PBCH \cite{3GPP38.212}, a 32-bit message is first encoded with a 24-bit CRC, then encoded with a (512, 56) polar encoder, and finally the first 352 bits are repeated to arrive at a final blocklength of 864.  We refer to a polar code concatenated with CRC as a CRC-polar code.

An inner code such as a polar code concatenated with an outer CRC may be decoded using CRC-aided list decoding.  With CRC-aided list decoding, two  different failure mechanisms produce decoding failures. One failure mechanism is that none of the $L$ paths in the list  pass the CRC check.  We refer to those decoding failures as {\it erasures}.
The second failure mechanism occurs when the path that passes the CRC check  is incorrect.
We refer to those decoding failures as 
{\it undetected frame errors}. The overall total failure rate including both 
type of errors is referred as the 
{\it total failure rate} (TFR). In \cite{KingICC}, it is shown that using a shorter CRC of 11 or 12 bits rather than 24 bits significantly reduces the TFR of the CRC-polar code.

An alternative to CRC-polar for the communication of short messages is the concatenation of a CRC with a tail biting convolutional code (TBCC) \cite{Ma1986}. For the resulting CRC-TBCC, CRC-aided list decoding may be accomplished with a list Viterbi algorithm (LVA) decoder \cite{Seshadri1994}.  CRC-aided list decoding of TBCCs has recently been shown to perform very well at short blocklengths, approaching and even surpassing the random coding union (RCU) bound \cite{Yang2022}, \cite{YangTBCC}.  

In \cite{KingICC}, an optimized CRC-TBCC provided a lower TFR than a similarly optimized CRC-polar code for the 5G PBCH 32-bit message.  The comparison in \cite{KingICC} suggests that CRC-TBCCs are a better solution than CRC-polar codes for the 5G standard, but the analysis is incomplete. For example, the code rates are not identical, the CRC for the polar code was not necessarily optimal, and the deficiencies of adaptive list decoding of a CRC-polar code were not fully understood.  This paper seeks to provide a careful comparison of CRC-polar and CRC-TBCC where the message size and codeword length are exactly the same, the best possible CRC is designed in each case, and decoder sub-optimality is fully considered.

Lou \emph{et al.} show the importance of designing CRC codes for specific convolutional codes \cite{Lou2015}.  They show that the frame error rate (FER) of the concatenated CRC-convolutional code is minimized by selecting the CRC that minimizes union bound on FER based on the distance spectrum of the CRC-convolutional code.  A CRC that minimizes this union bound for a specific convolutional code is known as \emph{distance-spectrum-optimal}, or DSO.  In this paper, we design a DSO 11-bit CRC for a specific TBCC, and we also design a DSO 11-bit CRC for the 5G Polar code by extending the ideas in \cite{Lou2015} to CRC-polar codes.

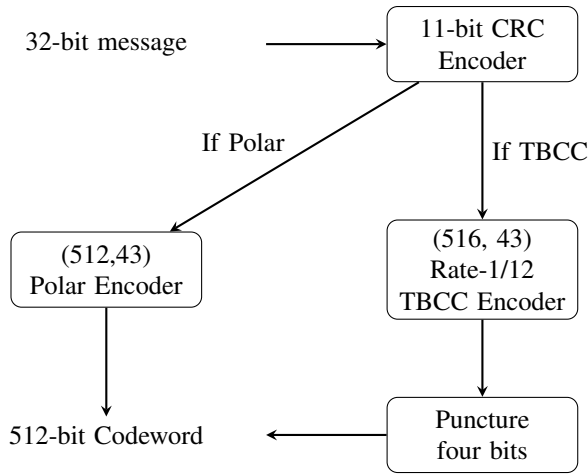
\begin{figure}

    \centering
    
    \tikzstyle{box} = [rectangle, rounded corners, minimum width=2.3cm, minimum height=1cm,text centered, text width=2.3cm, draw=black]
    \tikzstyle{no box} = [text centered, text width=4cm]
    \tikzstyle{arrow} = [thick,->,>=stealth]
    
    \begin{tikzpicture}[node distance=2cm]
    \node (box1) [no box] {32-bit message};
    
    \node (box2) [box, right of=box1, xshift=3cm] {11-bit CRC Encoder};
    \node (box3) [box, below of=box2, yshift=-1cm] {(516, 43) Rate-1/12 TBCC Encoder};
    \node (box6) [box, left of=box3, xshift=-3cm] {(512,43) Polar Encoder};
    \node (box4) [box, below of=box3, yshift=-0.2cm] {Puncture four bits};
    \node (box5) [no box, left of=box4, xshift=-3cm] {512-bit Codeword};
    
    \draw [arrow] (box1) -- (box2);
    \draw [arrow] (box2) -- node [anchor=west] {If TBCC} (box3);
    \draw [arrow] (box3) -- (box4);
    \draw [arrow] (box4) -- (box5);
    \draw [arrow] (box2) -- node [anchor=east, yshift=0.2cm] {If Polar} (box6);
    \draw [arrow] (box6) -- (box5);
    \end{tikzpicture}

    \caption{Encoding Scheme for CRC-polar and punctured CRC-TBCC codes.}
    \label{fig:Encoding}
\end{figure}

This paper provides CRC-polar and CRC-TBCC designs with equivalent message lengths, CRC lengths, and blocklengths so as to facilitate a fair comparison.  The designs avoid rate matching through bit repetition or puncturing as much as possible.  Since blocklengths that are powers of two are natural for polar codes, a blocklength of 512 bits is selected for our CRC-polar and CRC-TBCC designs.  The encoding schemes of the CRC-TBCC and CRC-polar are shown in Figure \ref{fig:Encoding}.

\subsection{Contributions}

This paper designs a new rate-1/12 TBCC, which we puncture to produce a (512, 43) TBCC that exactly matches the (512, 43) polar code. The paper then designs DSO 11-bit CRCs for the  CRC-polar and CRC-TBCC codes for message length and blocklength compatible with the 5G PBCH code.  We present the distance spectra of these two concatenated codes and analyze the consequences of the difference between them.  We show that our best CRC-TBCC design outperforms our best CRC-polar design, while also having a significantly faster decoder.

\subsection{Organization}

Sections \ref{sec:Polar_CRC} and \ref{sec:TBCC} give design criteria for a CRC-TBCC and CRC-polar codes.  These two sections identify the optimal code parameters and present a methodology for finding such optimal codes that minimize the union bound on FER.  Section \ref{sec:DistanceSpectra} provides the distance spectra of the CRC-polar and CRC-TBCC codes and explores the consequences of these distance spectra.  Lastly, section \ref{sec:Sim} provides simulation results for the error rates and decoder throughput.

\section{Polar Code Design}
\label{sec:Polar_CRC}
In the 5G New Radio (5G NR) 
technical specification~\cite{3GPP38.212}, 
six CRCs are proposed to concatenate with 
polar codes and LDPC codes for different 
message lengths.
Here, we list the six proposed CRCs in Table I. 
The generator polynomials of the 
CRCs are denoted in hexadecimal, 
where the high-order coefficients correspond to the 
most significant bits.
For example, the generator polynomial 
$x^{11}+x^{10}+x^9+x^5+1$ 
for CRC11 
in Table I
is denoted as 0xE21.

%
\begin{figure}[t]
    \centering
    \includegraphics[width=21pc]{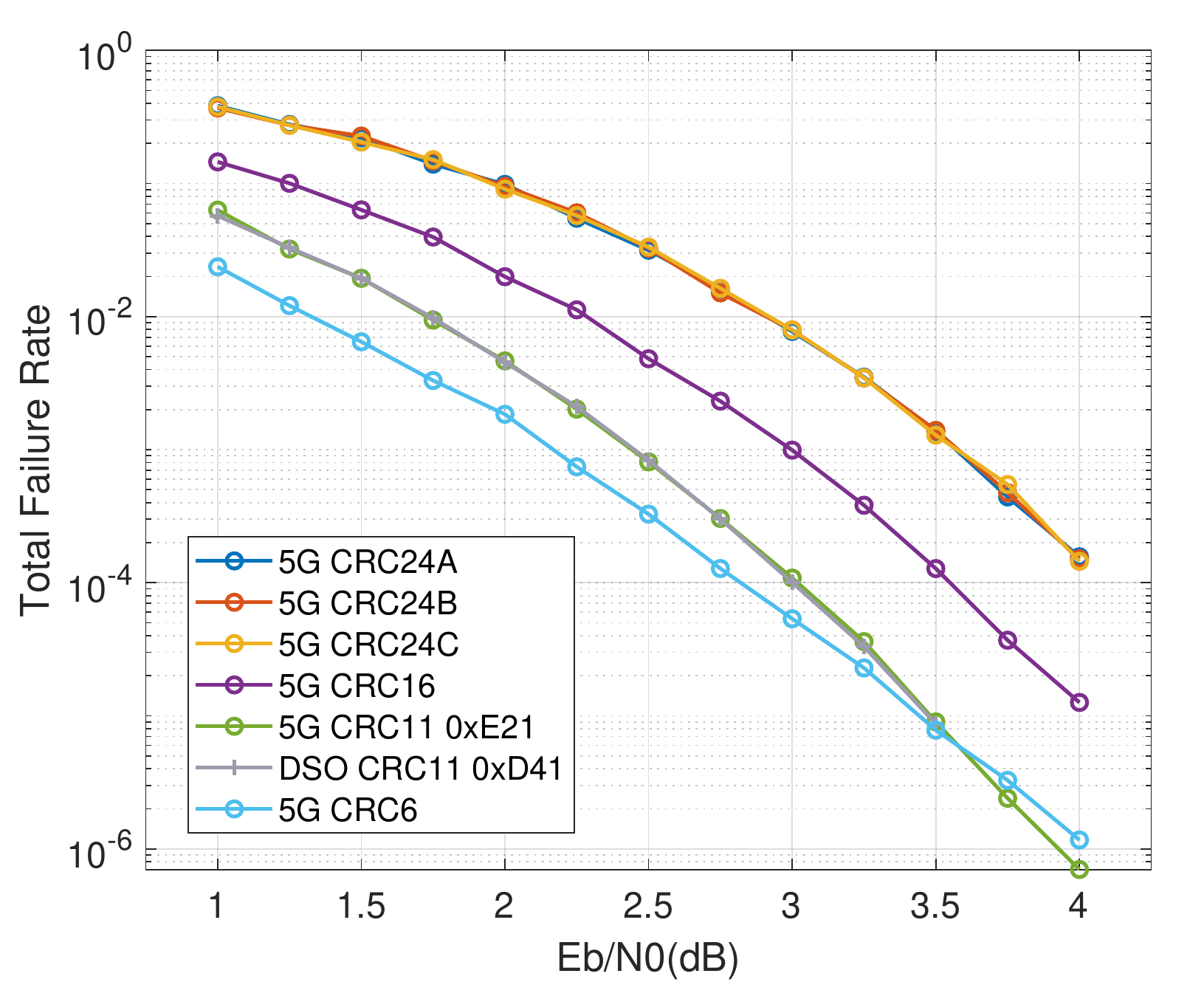}
    \caption{List decoding performance (list size $L=32$) 
    of length-512 5G polar codes with 32 message bits 
    and different CRCs}
    \label{fig:PolarCRCs}
\end{figure}

\begin{table}[t]
    \begin{center}
    \caption{CRCs proposed in the 
    5G NR technical specification
    \cite{3GPP38.212}}
    \begin{tabular}{|c|c|}
    \hline
         label &  generator polynomial  \\
    \hline
         CRC24A  & 0x1864CFB            \\
         CRC24B  & 0x1800063            \\
         CRC24C  & 0x1B2B117            \\
         CRC16   & 0x11021              \\
         CRC11   & 0xE21                \\
         CRC6    & 0x61                 \\
    \hline
    \end{tabular}
    \end{center}

    \label{tab:5G_CRCs}
\end{table}

For the PBCH polar code in 5G, 
the 32 message bits are first encoded 
with CRC24C listed in Table I. 
Then, both the 32 message bits 
and the appended 24 CRC bits 
are interleaved together and  
encoded using a (512,56) polar code. 
The frozen set of the polar code
is determined 
using 
the reliability sequence in 
the 5G standard~\cite{3GPP38.212}.
For a detailed description of this encoding process, 
we refer the readers to \cite{pillet2020list}.
Because they do not impact FER vs. Eb/No performance on a binary-input
AWGN channel, in our work we omit the rate matching, interleaving, and CRC parity distribution described in \cite{pillet2020list}.

%

%

On the receiver side, a common decoding 
approach for CRC-polar codes is
successive cancellation list (SCL)  decoding~\cite{ListSC}. 
In the list decoding process, $L$ candidate metric-ordered paths 
are generated, and the first path that passes the CRC check
will be selected as the decoding result. 
%

For a CRC-polar code with 32 message bits 
and $m$ CRC bits, 
$(32+m)$ synthesized bit channels of the inner polar code 
need to be unfrozen.
Regarding the length of the CRC, 
the following tradeoff can be observed.
By using a longer CRC, the probability that a random 
decoding path passes the CRC check at the 
end of the list decoding process will be smaller, 
resulting in an expected lower undetected frame error rate.
However, a longer CRC corresponds to more unfrozen
bit channels for the inner polar code. 
In this way, the total failure rate of the code might be 
damaged by introducing too many low reliability bit channels.
In light of this this tradeoff, 
we can explore what CRC length will minimize the total failure rate.   

As shown in \cite{KingICC}, 
the total failure rate 
can be significantly improved by replacing the 24-bit 
CRC24C with a shorter 11-bit CRC. 
For further investigation on this point, 
Fig.~\ref{fig:PolarCRCs} shows total failure rate vs. $E_b/N_0$ for all the CRCs (listed in Table I)
proposed in the 5G NR technical specification. In Fig.~\ref{fig:PolarCRCs}, the simulations are for the binary-input 
additive white Gaussian noise (AWGN) channel and the list size is set to the practically interesting value of $L = 32$.
%
Our result shows that among all the 5G CRCs, 
CRC6 has the best performance at low SNRs,
while the best performance at high SNRs 
is obtained by CRC11.
Following this result, 
if we limit our search space to CRC lengths specified 
in the 5G standard, 
CRCs with 11 bits achieve the best list decoding performance 
at high SNR.

However, CRC11 
is not known to be specifically designed 
together with the inner polar code. Before we compare with a CRC-TBCC, the next subsection designs the DSO 11-bit  CRC for the length-512 5G polar code with 32 message bits.
%

\subsection{Design of Distance Spectrum Optimal 11-bit CRC}
It is known that on AWGN channels, 
the maximum-likelihood (ML) decoding performance 
of binary linear codes 
is governed by their weight distribution, and
can be well approximated by the union bound. 
Here, we seek to design an 11-bit CRC that 
provides the best ML decoding performance at high SNRs.
For this purpose, 
we design a DSO 11-bit CRC. 


The union bound on FER based on the distance spectrum is given by

\begin{equation*}
    FER < \sum_{d = d_{min}}^n A(d) P_2(d).
\end{equation*}
Here, $A(d)$ is the number of codewords at weight $d$, $P_2(d)$ is the pairwise error probability of two codewords at distance $d$, and $n$ is the blocklength.  We wish to find the 11-bit CRC that minimizes this union bound of the CRC-polar concatenated code.  As SNR grows large, this minimization problem can be well approximated by maximizing the minimum distance $d_{min}$ and minimizing $A(d_{min})$.  We design an 11-bit CRC according to these criteria.

%
In the first step of our design procedure,
%
%
%
%
we use the algorithm in 
\cite{yao2021deterministic} to compute the entire
weight distribution of the (512, 43) inner polar code, 
whose frozen set is chosen according to the 
reliability sequence in the 5G standard~\cite{3GPP38.212}.
The partial weight distribution of this polar code 
for codewords with weight up to 128 is shown in 
Table II.

\begin{table}[h]
\begin{center}
\caption{Partial weight distribution of 
the (512,43) 5G polar code}
\begin{tabular}{|c|c|c|c|c|}
    \hline
    $d$   & 0 & 64 & 96 & 128 \\
    \hline
    $A(d)$ & 1 & 536 & 9600 & 496988\\
    \hline
\end{tabular}
\end{center}

\label{tab:wd_512_polar}
\end{table}

In the second step, 
we use the method in \cite{li2012adaptive} 
to obtain all codewords of weight 64, 
and all codewords of weight 96 for this polar code.
Consider the following experiment devised in
\cite{li2012adaptive}: 
Transmit the all-zero codeword through AWGN 
channels in the extremely high SNR regime, 
and decode the channel output using a list decoder. 
It is reasonable to expect that in this experiment, 
the list decoder will produce codewords of low weight. 
As the list size $L$ increases, 
since the decoder is forced to generate 
a list of size exactly $L$, 
more and more low-weight codewords emerge.
For the (512,43) 5G polar code,
by using list size $L = 32768$ in this experiment, 
we are able to obtain all 536 codewords of 
weight 64, and all 9600 codewords of weight 96.
The list size required for 
us to obtain all codewords of weight 128 is too large. 
Hence we stop this experiment at $L = 32768$.

In the third step, 
we go over all CRCs with 11 bits,
and check which one of them can eliminate 
most of those low weight codewords.
We consider all the 11-bit CRCs with generator polynomials 
in the form $x^{11}+\cdots+1$, such that the 
leading coefficient of $x^{11}$ and the constant term 
at the end are both fixed to be 1.
There are $2^{10}$ 11-bit CRCs in our search space, 
and we find out that 79 of them
can eliminate all codewords of weight 64 and weight 96 
for the (512,43) inner polar code. 

In the fourth step, 
we compute the complete weight distribution of 
the concatenated CRC-polar codes using all those 79 CRCs 
by the brute-force enumeration, and discover that the 
distance-spectrum-optimal CRC has generator polynomial 
0xD41.  This CRC-polar code has $d_{min} = 128$ and
$A(d_{min}) = 219$.
In Section \ref{sec:DistanceSpectra}, we 
present the distance spectra of the inner (512, 43) polar code 
and the CRC-polar code with 
CRC 0xD41.

%
%

\subsection{Adaptive List Decoding}

In this paper, we decode CRC-polar codes using a successive cancellation list (SCL) \cite{ChenSCL} decoder.  For large list sizes ($L > 64$), this requires significant computational cost to always generate a list of $L$ codewords for every message sent.  However, a large list size does tend to significantly improve TFR performance compared to smaller list sizes.  

We choose to implement an adaptive parallel list decoder to balance these effects, as in \cite{li2012adaptive}, \cite{KingICC}.  The adaptive list decoding algorithm is as follows.  We begin by running a parallel list decoder with an initial list size $L = L_{min}$.  If the decoder finds a codeword that passes the CRC check, it selects that codeword and terminates.  However, if no codeword on the list passes the CRC check, we double $L$ and run another parallel list decoder.  This continues until a valid codeword is found or the maximum list size $L_{max}$ is reached.  We will call such a decoder an $(L_{min}, L_{max})$ adaptive list decoder.

Unfortunately, the SCL decoder is not maximum likelihood, so changing the list size can change whether a codeword appears in the list or not.  As a result of this, the performance of the $(L_{min}, L_{max})$ adaptive SCL decoder performance is not identical to that of the nonadaptive SCL decoder with $L=L_{max}$. This behavior is exacerbated for smaller CRCs such as an 11-bit CRC because a valid but incorrect codeword is more likely to be chosen.  Care must be taken with selecting $L_{min}$ and $L_{max}$ of the adaptive SCL decoder.

\section{TBCC Code Design}
\label{sec:TBCC}

Recent results by Yang \emph{et al.} \cite{Yang2022} show that CRC-TBCCs with list decoding can approach and even surpass the RCU bound \cite{Polyanskiy}.  In this section we present this design procedure for a low rate CRC-TBCC to compare to the CRC-polar designed in the previous section.

A comparison of CRC-TBCCs with list decoding to the 5G PBCH CRC-polar code was performed in \cite{KingICC}, and in that comparison the CRC-TBCC code appears superior.  However, the TBCC design used in \cite{KingICC} was not a  completely fair comparison to the 5G CRC-polar code.  The TBCC in \cite{KingICC} was a memory-8, rate-1/5 convolutional code, borrowed from \cite{YangTBCC}.  With a 32-bit message and 11-bit CRC, the CRC-TBCC has a blocklength of 215 bits.  However, the 5G CRC-polar has a 512-bit blocklength before bit repetition, resulting in a significantly lower rate code than the CRC-TBCC.  This significant difference in blocklength is a disadvantage for the CRC-TBCC so that the comparison is not completely fair.  We solve this problem by designing a much lower rate TBCC to match the 512-bit blocklength of the CRC-polar code.

For this paper, we designed a memory-8, rate-1/12 TBCC which we concatenate with an 11-bit CRC, resulting in a (516, 32+11) CRC-TBCC.  We puncture four bits to arrive at a (512, 32+11) punctured CRC-TBCC, matching the rate of the 5G CRC-polar code.

\subsection{Design of TBCC and CRC Polynomials}

Similar to the design criteria for DSO CRCs, we can define a distance-spectrum-optimal TBCC as the set of convolutional code polynomials that minimizes the union bound on the distance spectrum of the TBCC.  We once again approximate this optimization by searching for the set of polynomials that maximize the minimum distance and minimize the number of codewords at the minimum distance.

A memory-8 binary convolutional code has $2^7 = 128$ possibilities for each polynomial.  Finding the optimal rate-1/12 convolutional code via brute force search requires searching ${128 \choose 12} \approx 2.4 \times 10^{16}$ polynomial combinations.  Even after eliminating equivalent polynomial combinations, the space of possible polynomial combinations is still far too large to exhaustively search through.  Thus, a non-exhaustive search is performed.  The search was performed by examining the initial weight spectra of thousands of randomly generated memory-8, rate-1/12 convolutional codes and storing the one(s) with the largest $d_{free}$ and the smallest $A(d_{free})$, $A(d_{free+1})$, $A(d_{free+2})$, and $A(d_{free+3})$.  From there, we selected the best polynomial combination found by this partial search.

Once the polynomials defining the TBCC  were selected, we then searched all $2^{10} = 1024$ possible 11-bit CRCs to find which CRC maximized $d_{min}$ of the CRC-TBCC concatenated code and minimized $A(d_{min})$.  We performed this search via an efficient CRC search algorithm for TBCCs described in \cite{Yang2020}.

\subsection{Optimal CRC-TBCC Parameters}

The best memory-8, rate-1/12, (516,43) TBCC that we found through our non-exhaustive search has generator polynomials \{533, 727, 765, 445, 715, 635, 563, 555, 737, 557, 677, 511\} in octal.  This TBCC has a minimum distance of $d_{free} = 75$, with a total of 86 codewords at weight 75.  Table III shows the weight distribution for the first few weights.

\begin{table}[h]
\begin{center}
\caption{Partial weight distribution of 
the (516,43) TBCC}
\begin{tabular}{|c|c|c|c|c|c|c|c|c|c|c|}
    \hline
    $d$   & 0 & 75 & 76 & 79 & 80 & 84 & 87 & 88 & 91 & 92 \\
    \hline
    $A(d)$ & 1 & 86 & 86 & 86 & 43 & 129 & 129 & 129 & 215 & 43\\
    \hline
\end{tabular}
\end{center}

\label{tab:wd_512_tbcc}
\end{table}

The 11-bit DSO CRC for this TBCC is 0xF69, where the most significant bit corresponds to the $x^{11}$ term of the polynomial.  The concatenated CRC-TBCC has a minimum distance of $d_{min} = 132$ and $A(d_{min}) = 37$.

We also must select four bits of the CRC-TBCC to puncture to match the length and rate of the CRC-polar code.  To select the optimal puncture pattern, we searched for the puncture positions that have the smallest effect on the minimum weight codewords of the CRC-TBCC; that is, we searched all weight-132 codewords for the positions that had the most 0's in that position.  This way, by puncturing these positions, there would be as small an impact on the minimum weight codewords as possible, thus maximizing the $d_{min}$ of the punctured codes.  We found that puncturing bit positions \{47, 60, 129, 504\} (where the codeword starts at bit position 0) has the least impact on these low-weight codewords.  This resulted in a punctured (512,32) CRC-TBCC code with $d_{min} = 130$ and $A(d_{min}) = 1$.

\subsection{Adaptive List Decoding}

We use an adaptive list Viterbi decoder for our CRC-TBCCs.  The algorithm is conceptually similar to that of the adaptive SCL decoder, but with a parallel LVA decoder \cite{Seshadri1994} instead of a parallel SCL decoder.  A critical difference is that the LVA decoder for list size $L$ gathers the $L$ trellis paths that have the $L$ highest likelihoods for each possible end state.  The decoder then combines the trellis paths of each end state into one list and orders them in terms of likelihood, and then selects the most likely codeword that passes both the tail-biting condition and the CRC check.  As the list size $L$ tends to $\infty$, the LVA decoder becomes the maximum likelihood decoder for CRC-TBCCs.

For each beginning/ending state, a parallel LVA decoder always finds the $L$ most likely codewords given the received noisy vector, ranked in order of most to least likely.  For example, if the correct codeword is the $k$th most likely codeword for a given end state and the given received noisy vector, then the codeword will not appear on the list if $L < k$, and it will appear at position $k$ if $L \ge k$.  Similarly, if there is an incorrect codeword with the same beginning/ending state as the correct codeword, the incorrect codeword passes the CRC check, and the incorrect codeword is more likely than the correct codeword, then the incorrect codeword will always appear before the correct one on the list.

A common practice for decoding TBCCs is to use the wrap around Viterbi algorithm (WAVA), which tends to improve performance over standard Viterbi decoding of TBCCs \cite{Shao2003}.  For our CRC-TBCC, we use a WAVA inspired algorithm where we perform a single pass through the trellis to initialize metrics before transitioning to adaptive LVA decoding.  Simulation results show that adaptive and nonadaptive LVA decoding with this single pass metric initialization have functionally identical TFR performance.

Results in \cite{Yang2022} show that as SNR increases, the average list rank of the decoded codeword converges very quickly to one.  For this reason, it makes sense to initialize our adaptive decoder with $L_{min}=1$.  This also implies that the adaptive decoder is significantly faster than the nonadaptive decoder.  If the correct codeword is also the most likely, the adaptive decoder will terminate after its first iteration with $L_{min}=1$, but the nonadaptive decoder will find all $L$ most likely codewords before selecting the correct one.  Simulation results in Section \ref{sec:Sim} confirm this fact.

\section{Distance Spectra Analysis}
\label{sec:DistanceSpectra}

\begin{figure}
    \centering
    \includegraphics[width=20pc]{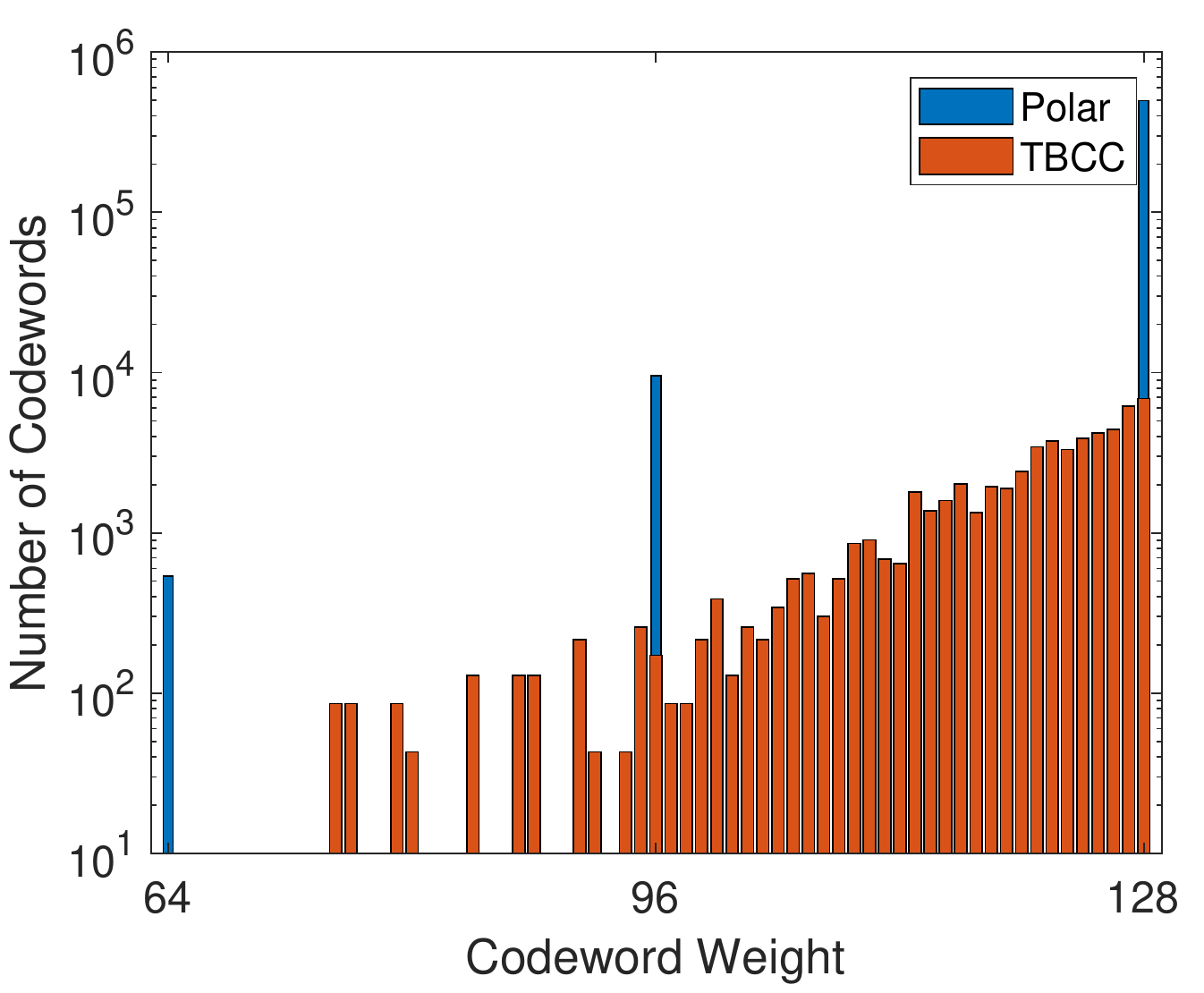}
    \caption{Partial weight distribution of (512,43) Polar and (516,43) TBCC.  Polar codewords are concentrated at specific weights, while TBCC codewords are more spread out between weights.}
    \label{fig:Partial_dist_no_crc}
\end{figure}

This section presents an in-depth comparison of the distance spectra for the polar, CRC-polar, TBCC, and CRC-TBCC codes that we have  designed  to make a fair comparison of (512, 32) block codes.

Since the codes we are working with have a relatively small number of codewords ($2^{32} \approx 4.3 \; \text{billion}$), we were able to compute the complete distance spectrum of the punctured CRC-TBCC and the CRC-polar codes.  Figure \ref{fig:Partial_dist_no_crc} shows the partial distance spectra of the (512, 43) Polar code and the (516, 43) convolutional code up to weight 128, plotted on a log scale.  This plot shows the distance spectra of both codes before the 11-bit CRC is applied to expurgate non-CRC-compliant codewords.  We can see that these two distance spectra are very different qualitatively.  The polar code has a very sparse distance spectrum, with codewords only appearing at weights that are multiples of 32, with large numbers of codewords concentrated at these distances.  By contrast, the TBCC distance spectrum is denser, with codewords appearing at every value of $d$, but with relatively low multiplicities at each individual distance.

Another property of the TBCC distance spectrum is that, at low weights, the number of codewords that appear at each weight is a multiple of the message length 43.  This stems from the fact that if you cyclically shift the message word of a TBCC by any amount, the codeword after convolutional encoding will also be a cyclic shift of the original codeword, so it will also have the same weight.  At low weights, codewords consist of a single error pattern surrounded by zeros, which means that low weight codewords are never cyclic.  This results in every cyclic shift of a low weight codeword being a unique codeword, and since there are 43 bits in the word before encoding, there are 43 unique cyclic shifts.  This breaks down at higher weights, where it is possible for a cyclic shift to produce a duplicate codeword.

\begin{figure}
    \centering
    \includegraphics[width=21pc]{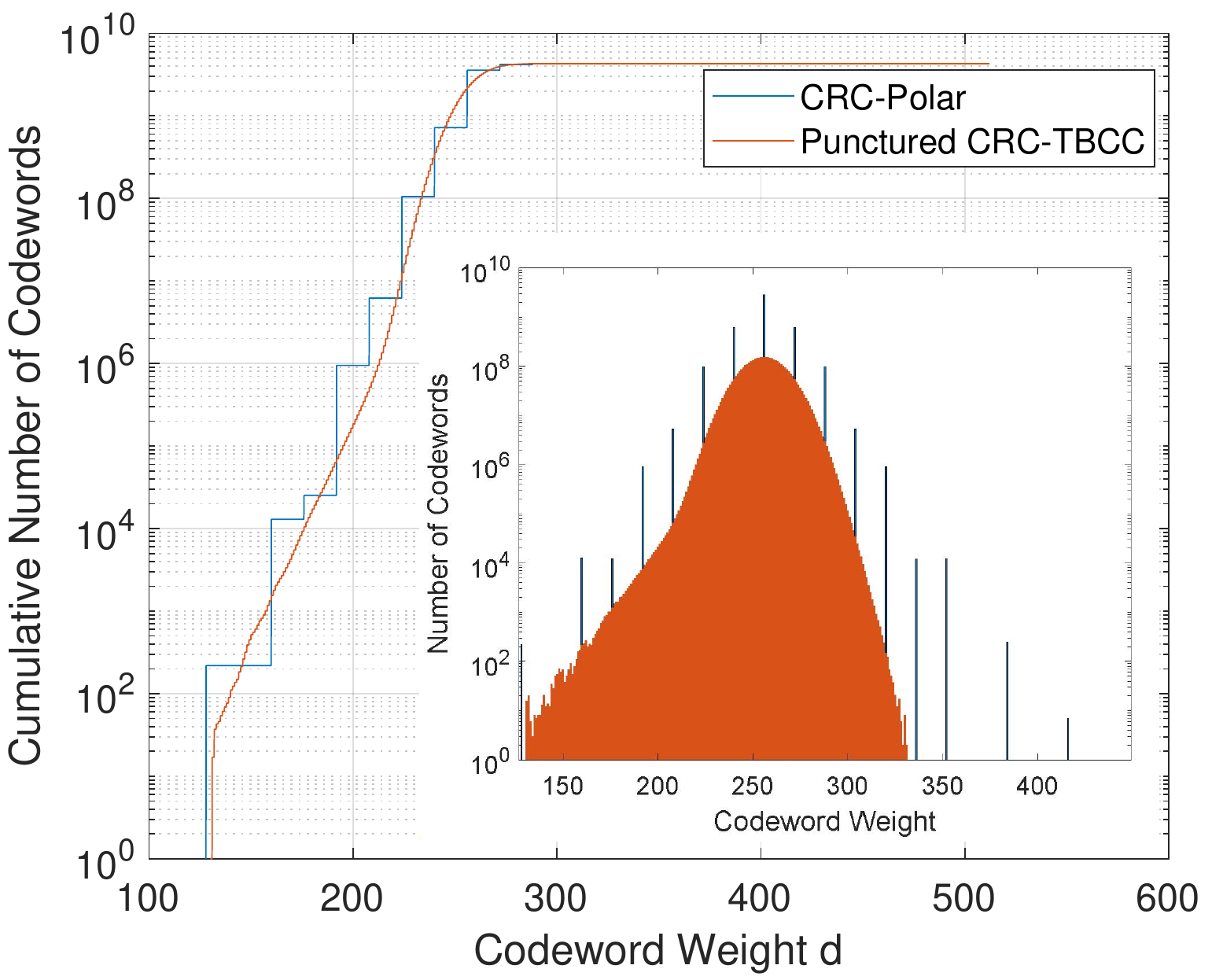}
    \caption{Distance spectra (inner plot) and cumulative distance spectra (outer plot) of CRC-polar and punctured CRC-TBCC.}
    \label{fig:Distance_Spectra_Log}
\end{figure}

Figure \ref{fig:Distance_Spectra_Log} shows the full distance spectra of the (512, 32) CRC-polar and the (512, 32) punctured CRC-TBCC codes on a log scale.  These codes have a very similar $d_{min}$, with the CRC-polar code having $d_{min} = 128$, and the punctured CRC-TBCC having $d_{min} = 130$.  Once again, the codewords of the CRC-polar code are concentrated at discrete weights (multiples of 16), while the CRC-TBCC distance spectrum forms a more continuous shape.

Figure \ref{fig:Distance_Spectra_Log} also shows the cumulative codeword distance spectra of each code.  That is, we plot the number of codewords with weight less than or equal to $d$ as a function of $d$.
We see that the cumulative codeword spectra of the two codes actually hug quite closely together.  The CRC-TBCC has fewer cumulative codewords at the smallest weights, and visual inspection seems to show that it tends to have fewer cumulative codewords than the CRC-polar at most weights.  However, the CRC-TBCC cumulative curve is not strictly below the cumulative curve of the CRC-polar code; there are several intervals where the CRC-polar code has fewer cumulative codewords.

\begin{figure}
    \centering
    \includegraphics[width=21pc]{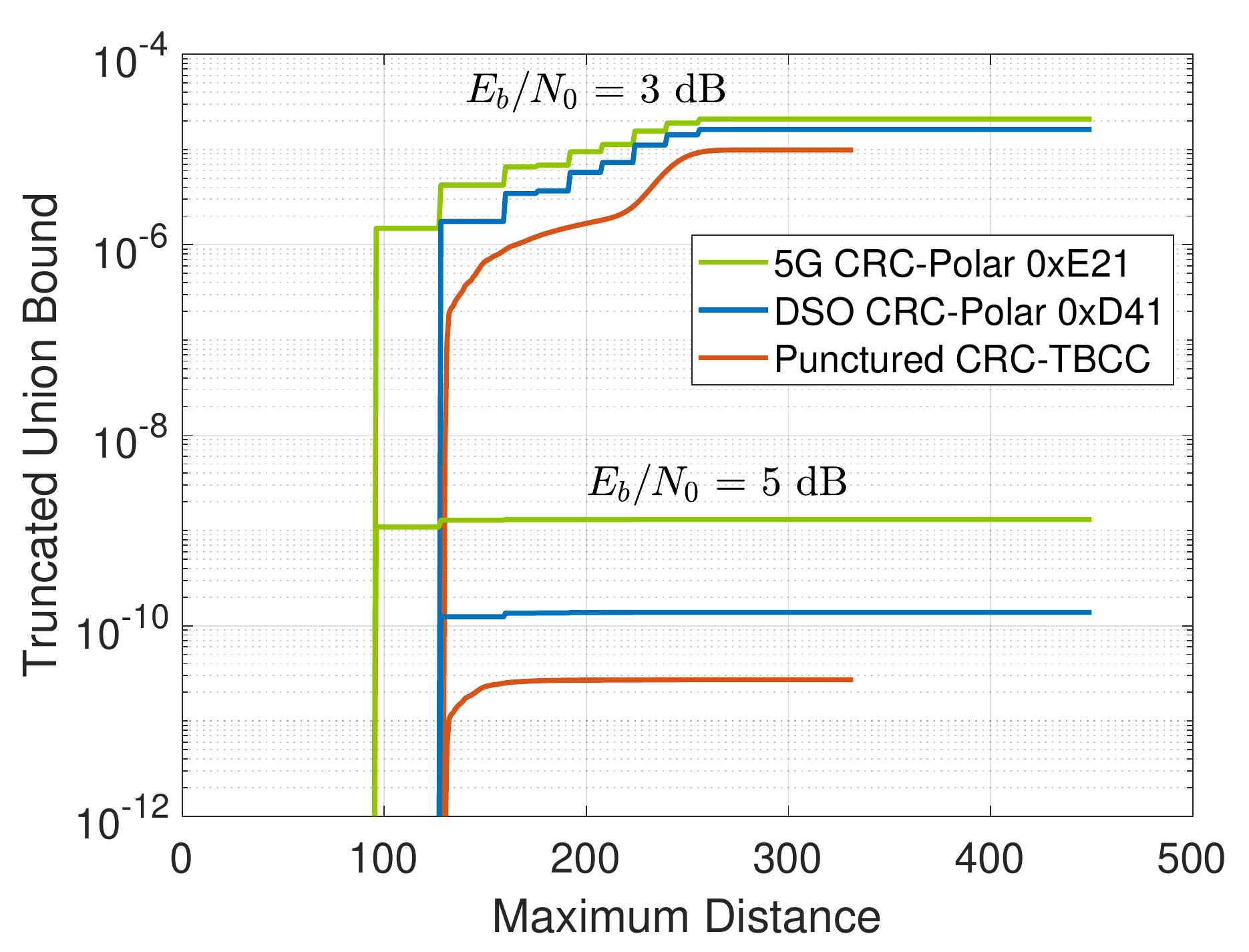}
    \caption{Truncated union bound of both codes vs. maximum distance.  The punctured CRC-TBCC has a strictly better truncated union bound than the DSO CRC-polar, which itself is better than the 5G CRC-polar code.  The CRC-TBCC has $d_{min} = 130$, the DSO CRC-polar has $d_{min} = 128$ and the 5G CRC-polar has $d_{min} = 96$.  As $E_b/N_0$ increases, the codewords at $d_{min}$ dominate the union bound.}
    \label{fig:Truncated_Union_Bound}
\end{figure}

To better understand the effects of the distance spectra, Figure \ref{fig:Truncated_Union_Bound} plots the truncated union bound up to weight $d$ vs. max weight $d$ for the punctured CRC-TBCC, the DSO CRC-polar, and the 5G CRC-polar codes.  We can observe a number of effects from Figure \ref{fig:Truncated_Union_Bound}.

Firstly, when comparing the CRC-TBCC and the DSO CRC-polar, the CRC-TBCC is the clear winner from the perspective of truncated union bound at both values of $E_b/N_0$.  The DSO CRC-polar code is unable to overcome the large union bound penalty that the 219 codewords at $d=128$ incurs at the start, and the CRC-TBCC has a strictly better truncated union bound at every maximum distance.  The same is true for the comparison between the DSO CRC-polar and the 5G CRC-polar code.  The 5G CRC-polar has a much worse $d_{min}$, and this manifests as a large union bound penalty when compared to the DSO CRC-polar.  From this comparison, we can see that replacing the 5G CRC with a DSO CRC for the polar code does noticeably improve the union bound.

Secondly, we notice a qualitative difference between the union bound curves for $E_b/N_0 = 3$ dB and $E_b/N_0 = 5$ dB.  At $E_b/N_0 = 3$ dB, we can see that the union bound keeps increasing by significant amounts until codeword weights of around 250.  This is roughly twice the $d_{min}$ of the CRC-TBCC and DSO CRC-polar codes, and $2.5 \times$ the $d_{min}$ of the 5G CRC-polar.  For this value of $E_b/N_0$, while the $d_{min}$ does impact the union bound, it is not the sole contributor to the final union bound value.

In contrast, when $E_b/N_0 = 5$ dB, the codewords at $d_{min}$ play a much more significant role in the value of the union bound.  For the CRC-TBCC, codewords stop mattering above a weight of around 150, and both CRC-polar codes have their union bounds determined almost entirely by their codewords at $d_{min}$.

As a consequence of the union bound becoming dominated by $d_{min}$ as $E_b/N_0$ increases, we notice that the gaps between the union bounds of the codes is larger at 5 dB than it is at 3 dB.  The CRC-TBCC has a larger $d_{min}$ and a $A(d_{min})$ than the other two codes, so its union bound performs significantly better as $E_b/N_0$ increases.  Figure \ref{fig:Truncated_Union_Bound} shows that our design algorithm for finding DSO CRCs by maximizing $d_{min}$ and minimizing $A(d_{min})$ becomes a better approximation as $E_b/N_0$ increases, and it also shows the importance of maximizing $d_{min}$ for codes operating at high $E_b/N_0$.

\section{Simulation Results}
\label{sec:Sim}

\begin{figure}
    \centering
    \includegraphics[width=21pc]{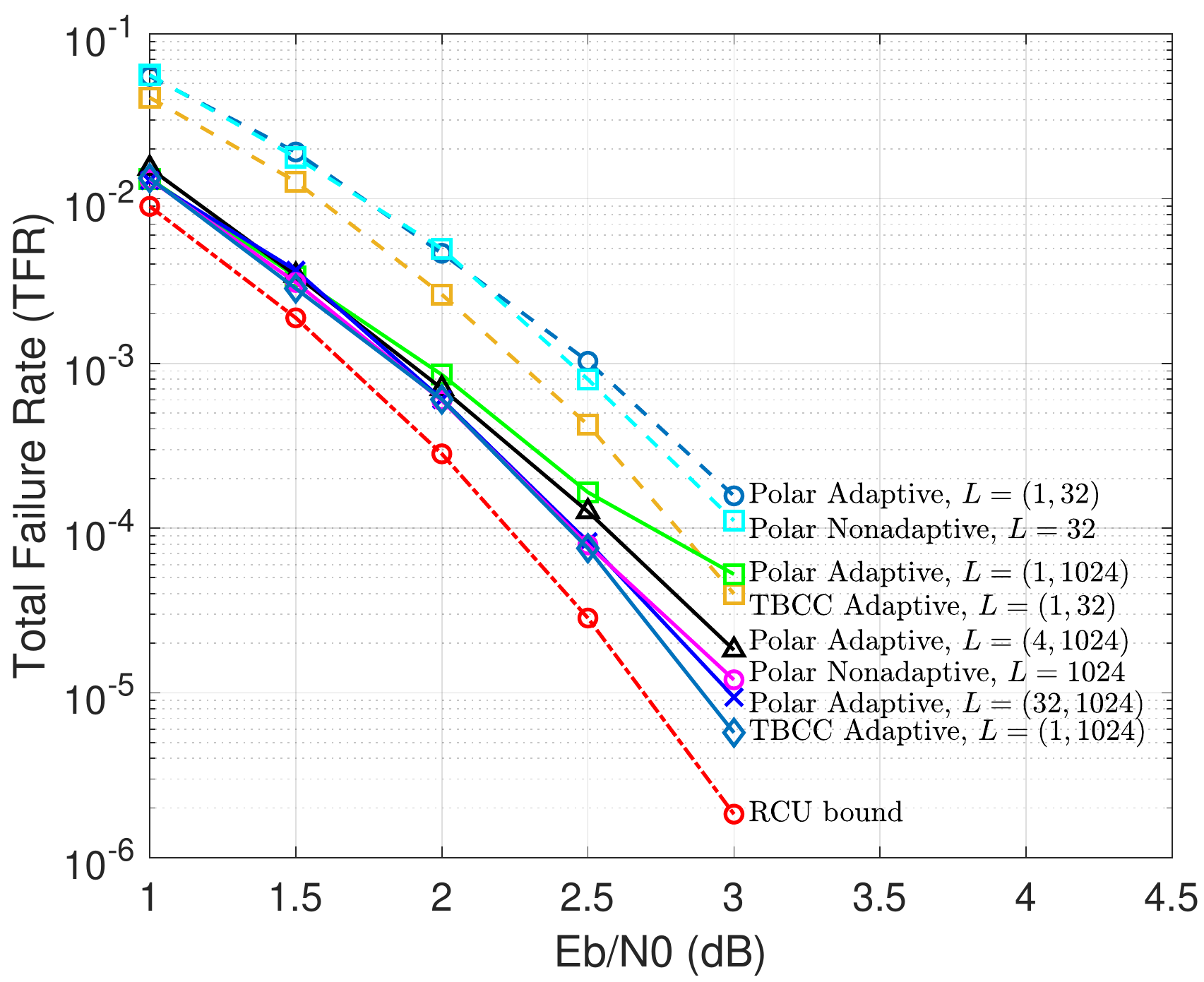}
    \caption{TFR vs. $E_b/N_0$ for all CRC-polar and CRC-TBCC codes.  The notation $L = (a,b)$ refers to an adaptive list decoder with $L_{min} = a$ and $L_{max} = b$. The CRC-TBCC has the best performance and is within 0.2 dB of the RCU bound.}
    \label{fig:CombinedPlot}
\end{figure}

We now present simulation results for our CRC-TBCC and CRC-polar designs.  We analyze the total failure rate performance of these codes, the trade off between list size and undetected error rate, and the decoding speed of our decoders in terms of throughput.  A more expansive analysis is available in \cite{KingThesis}.

\subsection{Total Failure Rate}

Figure \ref{fig:CombinedPlot} shows the TFR performance of the CRC-TBCCs and CRC-polar codes with various list-decoding schemes, plotted together with a saddlepoint approximation of the RCU bound \cite{Saddlepoint}.  We use maximum list sizes of $L_{max}=32$ and $L_{max}=1024$.  For the adaptive SCL decoder with $L_{max}=1024$, we also vary the minimum list size, $L_{min}$, between 1, 4, and 32.  For simplicity, we use the notation $L = (a,b)$ to refer to an adaptive list decoder with $L_{min} = a$ and $L_{max} = b$.

For $L_{max}=32$, the CRC-polar with nonadaptive SCL decoder has a slightly better TFR performance than its adaptive counterpart, gaining about 0.09 dB at TFR = $2 \times 10^{-4}$.  Both of these codes perform worse than the CRC-TBCC.

For $L_{max}=1024$, all codes have comparable TFR performances at low $E_b/N_0$.  For the adaptive CRC-polar curves, we see that changing $L_{min}$ has a significant impact on TFR performance at high $E_b/N_0$.  When $L_{min}=1$, there is a significant error floor starting at $E_b/N_0=2.5$ dB.  This error flooring effect is present in the comparisons in \cite{KingICC}, where an adaptive SCL decoder with $L_{min}=1$ was also used.  A SC decoder with list size 1 and 11-bit CRC turns out to have a high undetected error rate at high $E_b/N_0$, which explains this error floor.

\begin{figure}
    \centering
    \includegraphics[width=21pc]{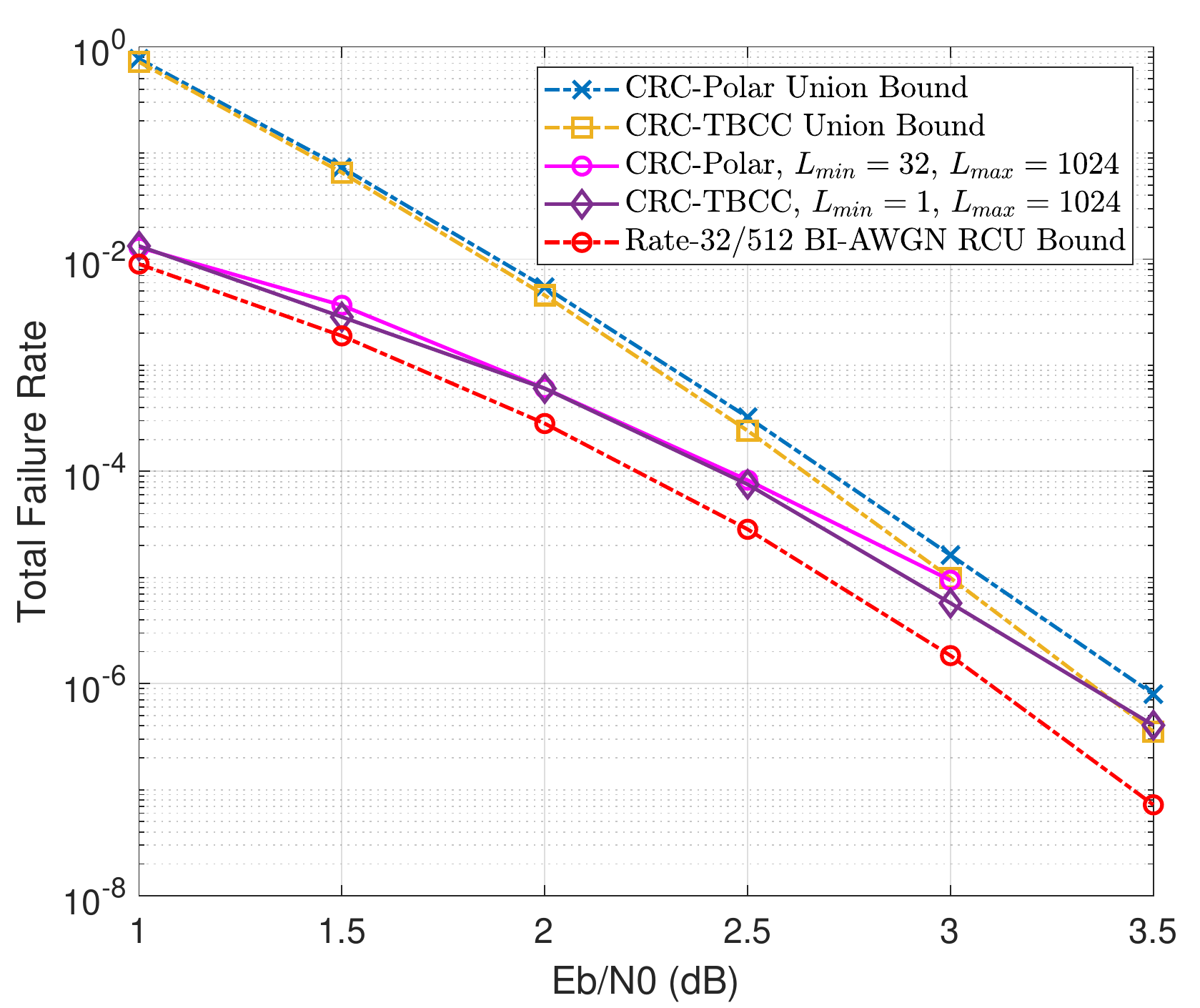}
    \caption{TFR curves of CRC-polar and CRC-TBCC and Union Bound curves.  The CRC-TBCC union bound separates from the CRC-polar union bound at high $E_b/N_0$, performing a little better. This effect is also seen in the TFR curves of the codes as they converge on union bound.}
    \label{fig:Union_Bound_Sim}
\end{figure}

As we increase $L_{min}$, we see that the performance of the adaptive CRC-polar improves until it is about equal to the performance of the nonadaptive CRC-polar when $L_{min} = 32$.  Due to the extremely slow speed of the nonadaptive SCL decoder with $L=1024$ (shown later) and the low target TFR, the 3 dB point of the nonadaptive CRC-polar curve  was computed with a relatively small sample size of error events (100).  As such, this data point is not completely reliable, which explains why the adaptive CRC-polar with $L_{min}=32$ is slightly better.

Once again, the best performing code for $L_{max}=1024$ is the CRC-TBCC, outperforming the best CRC-polar by nearly a factor of two at $E_b/N_0 = 3$ dB.  The CRC-TBCC TFR curve is about 0.2 dB away from the RCU bound.

In the previous section, we found the complete distance spectrum of the CRC-polar and CRC-TBCC codes.  With this information, we can plot the union bound of these codes against the simulation results.  Figure \ref{fig:Union_Bound_Sim} shows the union bound and the TFR curves for the best performing CRC-polar and CRC-TBCC with $L_{max}=1024$.  We can see that at high $E_b/N_0$, both codes hug very closely to the union bound.  Also, the CRC-TBCC has a slightly better union bound curve at high $E_b/N_0$, following the results from Figure \ref{fig:Truncated_Union_Bound}.  We expect the TFR curves to converge on the union bound as $E_b/N_0$ increases, meaning the CRC-TBCC will continue to outperform the CRC-polar.

\begin{figure}
    \centering
    \includegraphics[width=21pc]{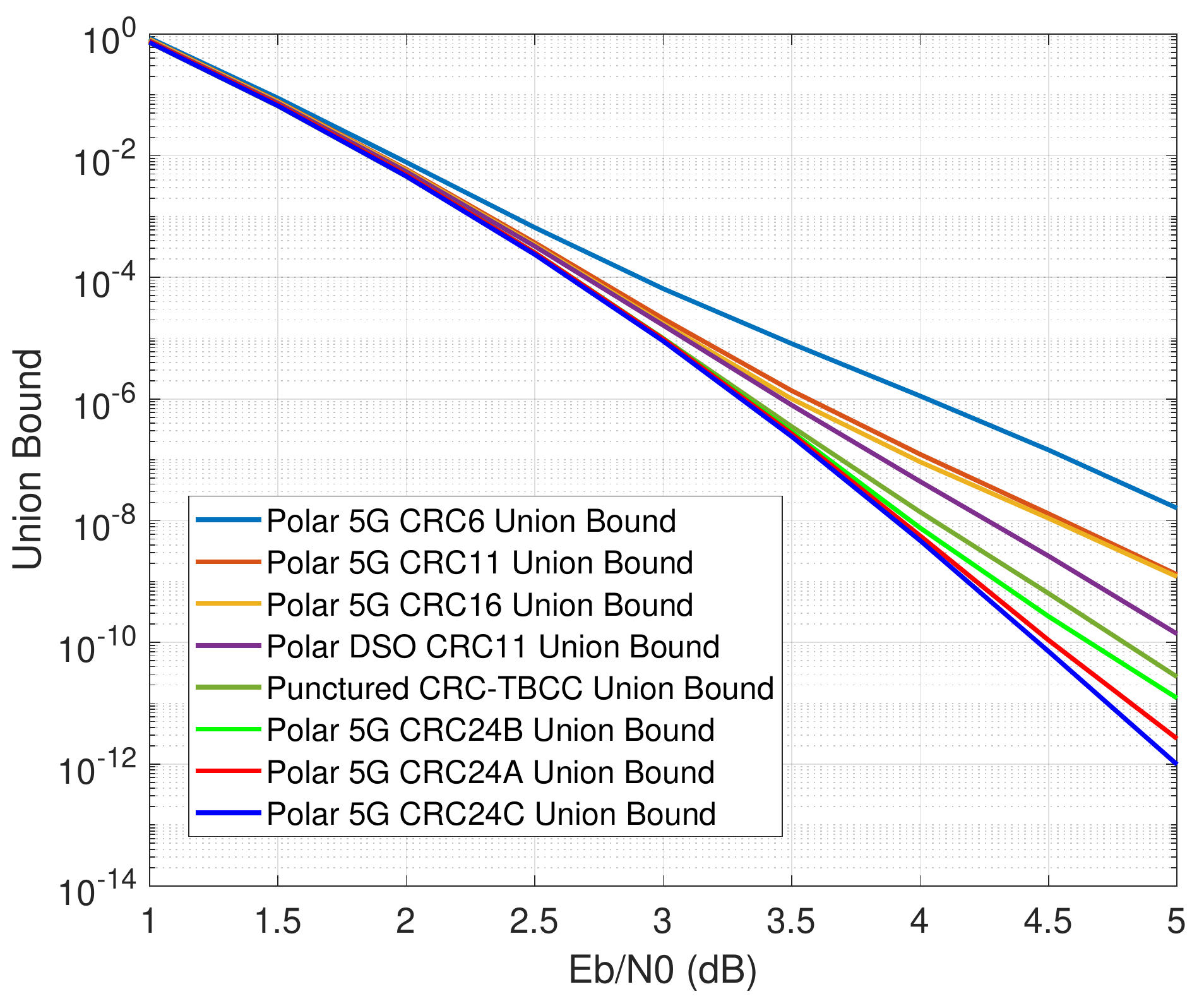}
    \caption{Union Bound curves for all 5G CRC-polar codes, as well as the 11-bit DSO CRC-polar and the CRC-TBCC.  Union bound tends to improve as CRC length improves, but the DSO CRC-polar and CRC-TBCC union bounds perform significantly better than the 5G CRC11 polar code, showing the importance of designing DSO CRCs.}
    \label{fig:Union_Bounds}
\end{figure}

Given how tightly our CRC-TBCC and CRC-polar codes hug the union bounds, we can get an idea for how well a code can do by looking at the union bound.  We decide to plot the union bounds of CRC-polar codes with every CRC from the 5G standard in Table I.  These union bounds are shown in Figure \ref{fig:Union_Bounds}, along with the union bounds of the punctured CRC-TBCC and our 11-bit DSO CRC-polar code.

Firstly, we can see that at high SNR, larger CRCs result in lower frame error rates according to union bounds, with the CRC6 polar code performing the worst, and the CRC24C polar code performing the best.  However, the simulation results in Figure \ref{fig:PolarCRCs} shows the the 24-bit CRC polar codes performs significantly worse than a CRC-polar with 11-bit CRC in total failure rate for a practical fixed list size of 32.  

In practice, the problem with a CRC-polar code with a 24-bit CRC is that the list size necessary to achieve good TFR performance is quite large.  In all of our simulations of CRC-polar codes with a 24-bit CRC, up to a maximum list size of 1024, we have never found an undetected error.  This indicates every single failure found has been a result of the maximum list size being too small.  Our 11-bit DSO CRC-polar, in contrast, has a large proportion of its failures as undetected errors, upwards of 80\% for high $E_b/N_0$.  As such, its TFR curve converges closely to its union bound curve.

From these results, in order for a code to achieve performance comparable to or better than the union bound, we expect that the maximum list size must be sufficiently large enough that most failures are undetected.  This intuitively makes sense, as the calculation of the union bound assumes all failures are undetected.  We conjecture that for a code with an $m$-bit CRC, a maximum list size of around $2^m$ is necessary for this union bound matching performance, motivated by simulation results in this paper and \cite{KingICC}, and by results in \cite{Yang2022} showing the expected list rank of CRC-convolutional codes converges to $2^m$ at low SNR.

Figure \ref{fig:Union_Bounds} also demonstrates the importance of designing DSO CRCs.  We can see that the DSO CRC-polar union bound outperforms the 5G CRC11 union bound by roughly a decade at $E_b/N_0 = 5$ dB, and even outperforms the union bound for CRC16.  This demonstrates that DSO CRC design becomes very important in the high SNR/low TFR region.  In addition, the union bound for the CRC-TBCC performs nearly as well as one of the 24-bit CRC-polar codes at $E_b/N_0 = 3$ dB.

\begin{figure}
    \centering
    \includegraphics[width=21pc]{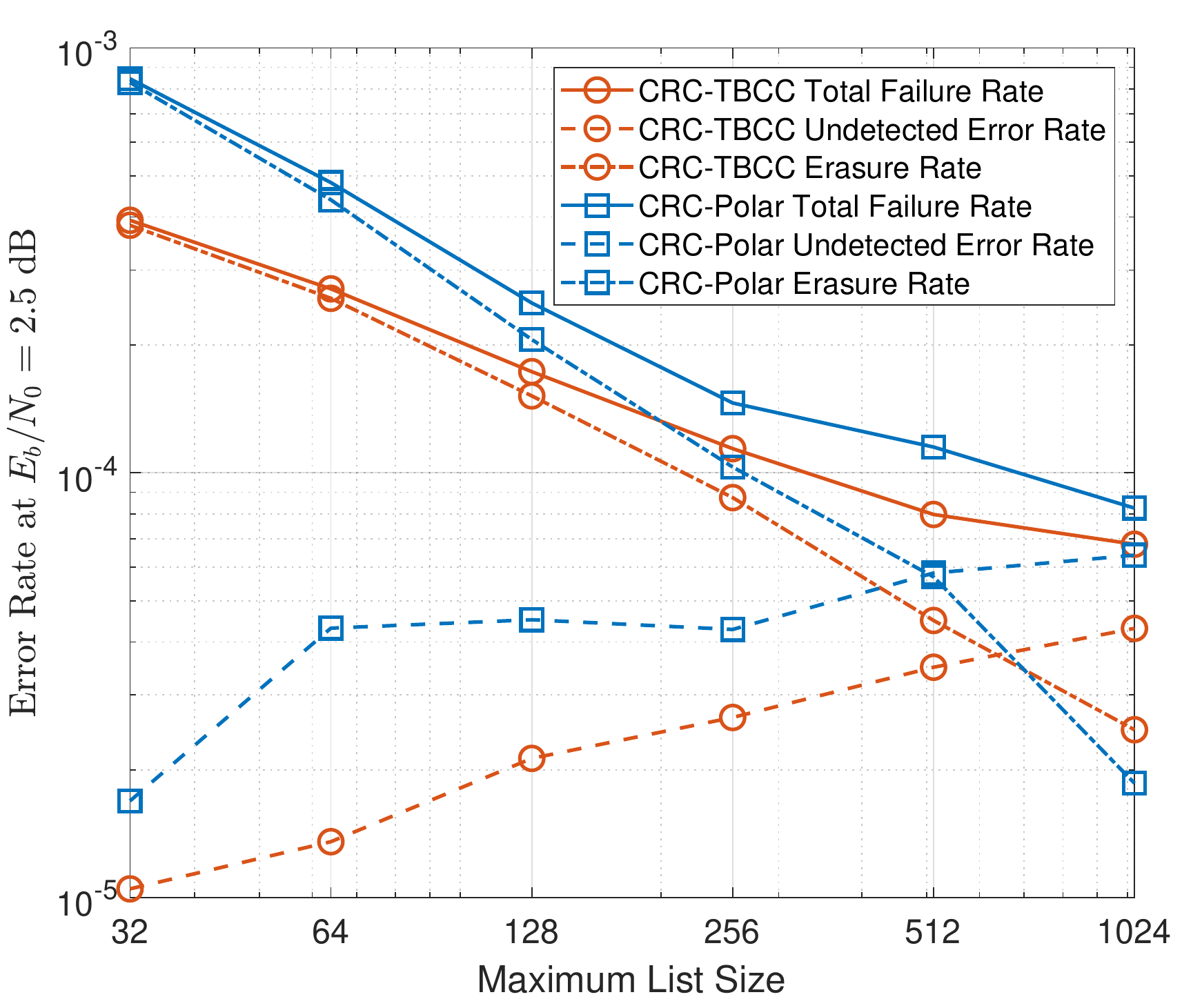}
    \caption{TFR, undetected error rate, and erasure rate curves as a function of maximum list size of adaptive decoder at $E_b/N_0 = 2.5$ dB.  Blue curves with square markers are $L = (32,L_{max})$ CRC-polar, and orange curves with circle markers are $L = (1, L_{max})$ CRC-TBCC.}
    \label{fig:Era}
\end{figure}

\subsection{Undetected Error Rate}

So far we have focused on the TFR performance of these codes and how different list sizes affects the TFR.  We will now explore the undetected error rate of these codes and its relation to TFR.

Figure \ref{fig:Era} plots the TFR, undetected error rate, and erasure rate of the $L_{min} = 1$ CRC-TBCC and the $L_{min} = 32$ CRC-polar codes at $E_b/N_0 = 2.5$ dB against maximum list size.  When $L_{max}$ is small, almost all errors are erasures, implying that $L_{max}$ is not large enough for optimal performance.  We confirm this when we increase $L_{max}$ and see that TFR performance improves significantly.  At large $L_{max}$, almost all errors are undetected, but even at $L_{max} = 1024$ that data imply that we could still improve TFR further by increasing $L_{max}$.

These data also suggest that an 11-bit CRC may be too short for applications where minimizing undetected errors is very important.  For these situations, a longer CRC will improve undetected error rate, but at the cost of TFR performance.  For example, the 24-bit CRC in the 5G standard has significantly worse TFR performance than an 11-bit CRC \cite{KingICC}, but its undetected error rate is substantially lower.  In fact, for a simulation of 300 error events, no undetected error events were recorded for the 24-bit CRC.  Alternatively, concatenating a second CRC will also substantially improve undetected error performance.

\subsection{Decoding Complexity}

\begin{figure}
    \centering
    \includegraphics[width=21pc]{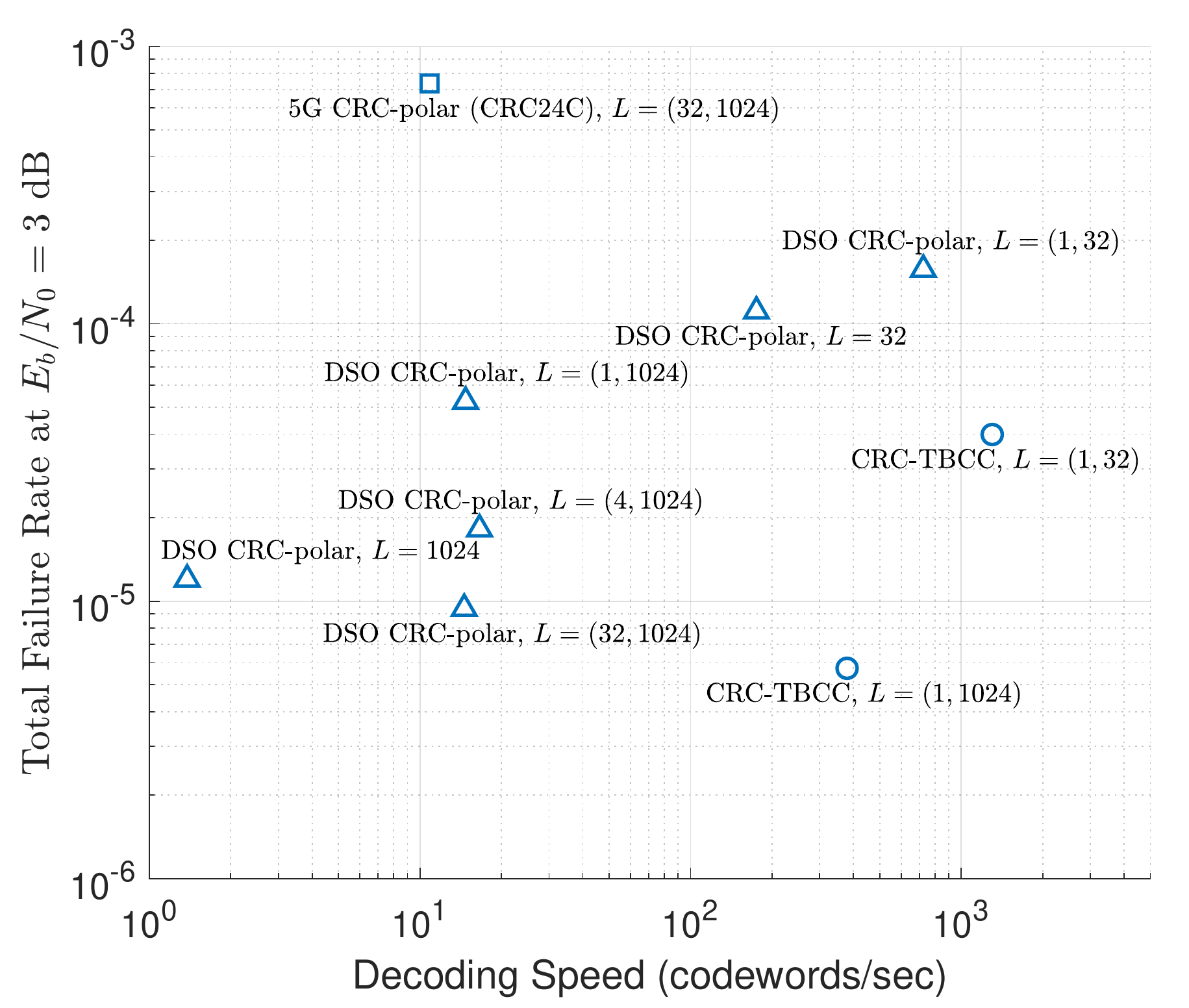}
    \caption{TFR at $E_b/N_0 = 3$ dB vs. decoded codewords per second for each decoder.  The $L_{max} = 1024$ LVA decoder for CRC-TBCC has the best TFR performance and significantly larger codeword throughput than the SCL decoders for CRC-polar.}
    \label{fig:Throughput}
\end{figure}

Finally, we present the decoding complexity  for each decoder.  Figure \ref{fig:Throughput} shows TFR performance at $E_b/N_0 = 3$ dB plotted against decoding run time\footnote{All simulations were performed on a System76 Galaga Pro Ubuntu laptop with an Intel Core i5-10210U CPU @1.6GHz x 8 Processor.}.

For both $L_{max} = 32$ and $L_{max} = 1024$ decoders, the CRC-TBCC with LVA decoder has both the fastest decoding speed and the best TFR performance.  In fact, for $L_{max} = 1024$, the LVA decoder is well over 10 times faster its SCL counterparts.

We also see that for the $L_{max} = 1024$ SCL decoders, increasing the minimum list size has negligible impact on decoding speed, but significant impact on TFR performance.  This is because $L_{min}$ is still small enough that the increase in complexity is small.  If we were to increase $L_{min}$ further, we would expect decoding complexity to converge toward the nonadaptive SCL decoder.

We also include the TFR and throughput of the 24-bit CRC-polar code used in \cite{KingICC}.  Figure \ref{fig:Union_Bounds} suggests that under optimal decoding this should outperform the 11-bit CRC-polar, but as discussed in Section V.A, the maximum list size of 1024 is not large enough for this result.  Instead, for $L = (32,1024)$, the 24-bit CRC-polar has similar decoding speed but significantly worse TFR performance compared to the 11-bit DSO CRC-polar.

\section{Conclusion}

In this paper we presented design methods for optimal low-rate CRC-polar and CRC-TBCC codes compatible with the 5G PBCH coding standard.  We designed these codes to have equivalent rate, blocklength, and CRC length in order to make the comparison between them as fair as possible.  We then did a direct comparison of these two codes, analyzing the properties of their distance spectra and comparing their error rates under simulation.  We found that, compared to the CRC-polar, the CRC-TBCC has a superior distance spectrum, a faster decoder, and better TFR performance.

We use a blocklength of 512 for our codes in this paper, which matches the blocklength of the polar encoder in the 5G PBCH code.  However, the 5G PBCH polar code actually has a blocklength of 864, where the first 352 bits of the 512-bit polar code are repeated.  We decided to ignore this repetition in this paper since repetition generates no improvement in TFR performance when compared against $E_b/N_0$ \cite{KingICC}.  However, a rate-1/20 TBCC with a 32-bit message and 11-bit CRC has a blocklength of 860 bits, which is very close to the 864-bit blocklength of the PBCH without repetition.  A well designed rate-1/20 CRC-TBCC could outperform the rate-1/12 CRC-TBCC we present in this paper and fit even better into the 5G standard.  This is an area of future interest.

In 2019, Ar{\i}kan presented an improvement on his polar codes, named \emph{polarization-adjusted-convolutional} (PAC) codes \cite{ArikanPAC}.  These codes have been shown to perform very well at short blocklengths, outperforming CRC-polar codes \cite{YaoPAC}.  We are interested in the performance of CRC-aided list decoding of PAC codes at these very low rates.

\section{Acknowledgements}

We thank Prof. Dariush Divsalar for the discussions we had on comparing CRC-TBCC and CRC-polar codes, and we credit Prof. Divsalar with the suggestion to lower the rate of our convolutional code design.  We also thank Dr. Hengjie Yang for his help with implementing the DSO CRC search algorithm for TBCCs and for his support and guidance throughout the writing of this paper.  Hanwen Yao would like to acknowledge the support and guidance of Prof. Alexander Vardy and Prof. Paul Siegel.

\bibliography{IEEEabrv, paper_ref}

\clearpage

\end{document}